\documentclass{article}
\usepackage{amsmath,amssymb,bbm}
\usepackage{mathrsfs}
\usepackage{cite}

\newcommand{\wt}[1]{\widetilde{#1}}

\begin{document}
\title{On the Physical Interpretation of the Dirac Wavefunction II: The Massive Dirac Field}
\author{Anastasios Y. Papaioannou \\ \tt{tasosp@gmail.com}}

\maketitle

\begin{abstract}
Using the language of the Geometric Algebra, we recast the massive Dirac bispinor as a set of Lorentz scalar, vector, bivector, pseudovector, and pseudoscalar fields that obey a generalized form of Maxwell's equations of electromagnetism. This field-based formulation requires careful distinction between geometric and non-geometric implementations of the imaginary unit scalar in the Dirac algebra. This distinction, which is obscured in conventional treatments, allows us to find alternative constructions of the field bilinears and a more natural interpretation of the discrete $C$, $P$, and $T$ transformations.
\end{abstract}

\section{The massless Dirac equation}
\label{sec:massless}
We first review the use of the projection bispinor \cite{papaioannou} to derive the multivector-valued version of the massless Dirac equation
\begin{equation*}
\nabla \psi = 0,
\end{equation*}
where
\begin{equation*}
\nabla \equiv \partial_\mu \gamma^\mu.
\end{equation*}
The Dirac bispinor can be factorized into two terms,
\begin{equation}
\psi = \psi_M w,
\end{equation}
where the dynamical degrees of freedom are contained within the multivector field
\begin{equation*}
\psi_M = f \, 1 + v_\mu \gamma^\mu + \frac{1}{2!} F_{\mu \nu} \gamma^\mu \wedge \gamma^\nu + p_\mu I \gamma^\mu + g I,
\end{equation*}
$w$ is a fixed bispinor, and we have used notation familiar from the Geometric Algebra:
\begin{align*}
\gamma^\mu \wedge \gamma^\nu & \equiv \frac{1}{2} (\gamma^\mu \gamma^\nu - \gamma^\nu \gamma^\mu) \\
I & \equiv \gamma_0 \gamma_1 \gamma_2 \gamma_3 = -\gamma^0 \gamma^1 \gamma^2 \gamma^3.
\end{align*}
The multivector field $\psi_M$ contains sixteen real degrees of freedom: the scalar $f$, four vector components $v_\mu$, six bivector components $F_{\mu \nu}$, four pseudovector components $p_\mu$, and one pseudoscalar $g$. These field components are all real-valued; that is, $\psi_M$ is an element of the \textit{spacetime algebra}, namely, the Clifford algebra $Cl_{1,3}(\mathbb{R})$.

The sole purpose of the fixed bispinor $w$ is to project these dynamical degrees of freedom onto an abstract vector space. We choose $w$ to obey the relations
\begin{align*}
\gamma^0 w & = w \\
\gamma^2 \gamma^1 w & = i w,
\end{align*}
which in the Dirac basis gives it the form
\begin{equation}
w = \begin{pmatrix} 1 \\ 0 \\ 0 \\ 0 \end{pmatrix}.
\end{equation}
The massless Dirac equation, with eight real degrees of freedom,
\begin{equation*}
\nabla \psi_M w = 0
\end{equation*}
can be expanded into a multivector equation with sixteen real degrees of freedom:
\begin{equation} \label{eq:multivectormassless}
\nabla \psi_M = 0.
\end{equation}
By the nature of $w$, which projects $\gamma^0$ and $1$ onto the same spinor component, this derivation of the multivector equation from the spinor equation is not unique. We justify it, however, as being Lorentz invariant, with no mixing of, e.g., scalar and vector terms. In component form, eq. \ref{eq:multivectormassless} becomes
\begin{align} \label{eq:maxwell}
\partial^\alpha f + \partial_\beta F^{\beta \alpha} & = 0 \nonumber \\
\partial^\alpha g + \partial_\beta \mathscr{F}^{\beta \alpha} & = 0 \nonumber \\
\partial_\alpha v^\alpha & = 0 \nonumber \\
\partial_\alpha p^\alpha & = 0 \nonumber \\
F_v^{\alpha \beta} + F_p^{\alpha \beta} & = 0,
\end{align}
where
\begin{align*}
\mathscr{F}^{\alpha \beta} & \equiv \frac{1}{2!} \epsilon^{\alpha \beta \gamma \delta} F_{\gamma \delta} \\
F_v^{\alpha \beta} & \equiv \partial^\alpha v^\beta - \partial^\beta v^\alpha \\
F_p^{\alpha \beta} & \equiv \frac{1}{2!} \epsilon^{\alpha \beta \gamma \delta} (\partial_\gamma p_\delta - \partial_\delta p_\gamma).
\end{align*}
There is no mixing between even-grade (scalar, bivector, and pseudoscalar) and odd-grade (vector and pseudovector) fields. For simplicity, we will assume that the odd fields vanish, leaving eight degrees of freedom:
\begin{equation}
\psi_M = f \, 1 + \frac{1}{2!} F_{\mu \nu} \gamma^\mu \wedge \gamma^\nu + g I,
\end{equation}
or, in spinorial form,
\begin{equation}
\psi = \begin{pmatrix} f - i F_{1 2} \\ - i F_{2 3} + F_{3 1} \\ i g - F_{3 0} \\ - F_{1 0} - i F_{2 0} \end{pmatrix}.
\end{equation}

\subsection{Lorentz transformations}
In the spacetime algebra, we interpret the gamma matrices $\{ \gamma^\mu \}$ as matrix representations of physical basis vectors. Under a Lorentz transformation between two reference frames, these basis vectors transform as
\begin{equation*}
\gamma^\mu \to \gamma'^\mu = R \, \gamma^\mu \, \wt{R},
\end{equation*}
where
\begin{equation*}
R = \exp(- \frac{1}{4} \omega_{\mu \nu} \gamma^\mu \wedge \gamma^\nu)
\end{equation*}
is the half-angle / half-rapidity Lorentz transformation operator, and $\wt{R}$ is the multivector \textit{reverse} of $R$, in which all constituent vectors in $R$ are reverse-ordered, i.e.,
\begin{equation*}
(\gamma^\mu \wedge \gamma^\nu)^\sim = \gamma^\nu \wedge \gamma^\mu = - (\gamma^\mu \wedge \gamma^\nu).
\end{equation*}
The reversed operator
\begin{equation*}
\wt{R} = \exp(+ \frac{1}{4} \omega_{\mu \nu} \gamma^\mu \wedge \gamma^\nu)
\end{equation*}
is thus the multiplicative inverse of $R$, with $\omega_{\mu \nu} \to -\omega_{\mu \nu}$. Higher-grade elements inherit their Lorentz transformation properties from the vectors comprising them:
\begin{align*}
\gamma'^{\mu_1} \gamma'^{\mu_2} \cdots \gamma'^{\mu_n} & = (R \gamma^{\mu_1} \wt{R}) (R \gamma^{\mu_2} \wt{R}) \cdots (R \gamma^{\mu_n} \wt{R}) \\
& = R (\gamma^{\mu_1} \gamma^{\mu_2} \cdots \gamma^{\mu_n}) \wt{R}.
\end{align*}
That is, elements of all grades transform in the same manner:
\begin{equation*}
M \to R M \wt{R}.
\end{equation*}

To derive the corresponding expression for the transformation of a multivector component under a change of reference frame, we contract $\psi_M$ with the appropriate basis element in the new reference frame, e.g.,
\begin{equation*}
v'^\mu = \langle \psi_M \gamma'^\mu \rangle = \langle \psi_M (R \gamma^\mu \wt{R}) \rangle,
\end{equation*}
where $\langle M \rangle$ denotes the scalar term of the multivector $M$. By the cyclic property of products in a scalar term,
\begin{equation*}
\langle \psi_M (R \gamma^\mu \wt{R}) \rangle = \langle (\wt{R} \psi_M R) \gamma^\mu \rangle,
\end{equation*}
which is the contraction of the inverse-transformed multivector onto $\gamma^\mu$.
Similar rules hold for the other basis elements. Under a change of reference frame, then, we can write $\psi_M \to S \psi_M \wt{S}$, where the transformation operator is
\begin{equation*}
S = \wt{R} = \exp(+ \frac{1}{4} \omega_{\mu \nu} \gamma^\mu \wedge \gamma^\nu),
\end{equation*}
so that each term in $S \psi_M \wt{S}$ is the Lorentz-transformed component:
\begin{equation*}
S \psi_M \wt{S} = f' 1 + v'_\mu \gamma^\mu + \frac{1}{2!} F'_{\mu \nu} \gamma^\mu \wedge \gamma^\nu + p'_\mu I \gamma^\mu + g' I.
\end{equation*}
Given the one-sided transformation rule for the Dirac bispinor:
\begin{equation*}
\psi \to \psi' = S \psi,
\end{equation*}
we find the corresponding one-sided transformation rule also holds for the projection bispinor $w$:
\begin{equation*}
\psi_M w \to S \psi_M w = (S \psi_M \wt{S}) (S w).
\end{equation*}

\subsection{Bilinears}
The factorization of the Dirac bispinor allows us to construct bilinears solely in terms of multivectors. For a general multivector element $M$, the bilinear $\overline{\psi} M \psi$ takes the form
\begin{align*}
\overline{\psi} M \psi = (w^\dagger \psi_M^\dagger \gamma^0) M (\psi_M w) \\
= w^\dagger \gamma^0 \wt{\psi}_M M \psi_M w \\
= (\gamma^0 \wt{\psi}_M M \psi_M)_{1 1},
\end{align*}
where we have written the Hermitian conjugate in a basis-independent form using the reversal operation:
\begin{equation*}
\psi_M^\dagger = \gamma^0 \wt{\psi}_M \gamma^0.
\end{equation*}
In the Dirac basis, only $1$, $\gamma^0$, $\gamma^2 \gamma^1$, and $\gamma^2 \gamma^1 \gamma^0$ have non-zero $(1,1)$ components. The bilinear expression therefore projects out the terms proportional to these four basis elements:
\begin{equation}
\overline{\psi} M \psi = \langle (\gamma^0 \wt{\psi}_M M \psi_M) (1 + \gamma^0 - i \gamma^2 \gamma^1 - i \gamma^2 \gamma^1 \gamma^0) \rangle.
\end{equation}
For example, the vector bilinear $j^\mu$ for the massless Dirac field is
\begin{align*}
\overline{\psi} \gamma^\mu \psi & = \langle (\gamma^0 \wt{\psi}_M \gamma^\mu \psi_M) (1 + \gamma^0 - i \gamma^2 \gamma^1 - i \gamma^2 \gamma^1 \gamma^0) \rangle \\
& = \langle \psi_M (1 + \gamma^0 - i \gamma^2 \gamma^1 - i \gamma^2 \gamma^1 \gamma^0) \gamma^0 \wt{\psi}_M \gamma^\mu \rangle \\
& = \langle (\psi_M \gamma^0 \wt{\psi}_M) \gamma^\mu \rangle,
\end{align*}
which gives the $\gamma^\mu$ component of the vector quantity $j = \psi_M \gamma^0 \wt{\psi}_M$. In component form, we have
\begin{equation}
j = (f^2 + g^2 + \vec{E}^2 + \vec{B}^2) \gamma^0 - 2 (f \vec{E} + g \vec{B} + \vec{E} \times \vec{B})_i \gamma^i.
\end{equation}
To draw out the similarity to electromagnetism, we have written the bivector components as
\begin{align*}
E_i & = F^{i 0} \\
B_i & = - \frac{1}{2} \epsilon^{i j k} F^{j k}.
\end{align*}
Via similar methods, the pseudovector spin angular momentum density is
\begin{align*}
S^{0 i j} & = \frac{i}{2} \overline{\psi} \gamma^0 (\gamma^i \wedge \gamma^j) \psi \\
& = \frac{i}{2} w^\dagger \gamma^0 \wt{\psi}_M \gamma^0 (\gamma^i \wedge \gamma^j) \psi_M w \\
& = \frac{1}{2} \langle \gamma^0 \wt{\psi}_M \gamma^0 (\gamma^i \wedge \gamma^j) \psi_M i (1 + \gamma^0 - i \gamma^2 \gamma^1 - i \gamma^2 \gamma^1 \gamma^0) \rangle \\
& = \frac{1}{2} \langle \gamma^0 \wt{\psi}_M \gamma^0 (\gamma^i \wedge \gamma^j) \psi_M (\gamma^2 \gamma^1 + \gamma^2 \gamma^1 \gamma^0 + i 1 + i \gamma^0) \rangle \\
& = \frac{1}{2} \langle (\psi_M \gamma^2 \gamma^1 \gamma^0 \wt{\psi}_M) \gamma^0 (\gamma^i \wedge \gamma^j) \rangle.
\end{align*}
This gives the $(0ij)$ component of the pseudovector quantity
\begin{equation}
S = \frac{1}{2} \psi_M \gamma^2 \gamma^1 \gamma^0 \wt{\psi}_M.
\end{equation}
For $i = 1$, $j = 2$, we have the ``$z$-component'' of the spin, which in component form is:
\begin{equation*}
S^{0 1 2} = \frac{1}{2} (f^2 + g^2 - \vec{E}^2 - \vec{B}^2 + 2 E^2_3 + 2 B^2_3).
\end{equation*}

Note that the factor of $i$ in the bilinear effectively serves as multiplication by $\gamma^2 \gamma^1$:
\begin{equation*}
i(1 + \gamma^0 - i \gamma^2 \gamma^1 - i \gamma^2 \gamma^1 \gamma^0) = \gamma^2 \gamma^1 (1 + \gamma^0 -i \gamma^2 \gamma^1 - i \gamma^2 \gamma^1 \gamma^0),
\end{equation*}
allowing us to implement factors of the imaginary unit scalar using an element of the real algebra $Cl_{1,3} (\mathbb{R})$:
\begin{equation*}
i \psi = i \psi_M w = \psi_M \gamma^2 \gamma^1 w.
\end{equation*}

\subsection{The field and operator interpretations}
We emphasize the distinction between our ``field interpretation'' of the multivector $\psi_M$ and the conventional ``operator interpretation'' in other treatments of the Dirac field in the spacetime algebra \cite{doran-lasenby}, \cite{hestenes-space-time-algebra}, \cite{hestenes-local-observables}. In the field interpretation, $\psi_M$ describes dynamic field degrees of freedom. The massless Dirac equation, for example, describes the dynamic behavior of one scalar, one pseudoscalar, and six bivector flux fields:
\begin{equation*}
\psi_M = f \, 1 + \frac{1}{2!} F_{\mu \nu} \gamma^\mu \wedge \gamma^\nu + g I,
\end{equation*}
which obey a generalized form of Maxwell's equations. The projection bispinor is merely a mathematical device that allows for simplified calculations within matrix representations of the underlying algebra, by projecting those degrees of freedom onto an abstract vector space.

In other formulations of the Dirac equation in the spacetime algebra, the multivector is instead treated as a dynamic transformation operator that acts upon a constant reference bispinor. In this operator interpretation, $\psi_M$ transforms the bispinor and other constant reference frame basis elements. Bilinears are interpreted as transformed basis elements, not as derived quantities that are quadratic in a more fundamental field quantity. In the field interpretation, the field is fundamental and the bilinears are derived quantities.

\section{The massive Dirac equation}
The Klein-Gordon equation for a real scalar field $f$ is
\begin{equation*}
\Box f = -{\omega_0}^2 f,
\end{equation*}
where $\Box \equiv \partial_{tt} - c^2 \vec{\nabla}^2$ and $\omega_0 = m c^2 / \hbar$. To derive the corresponding first-order equations, we first extend $f$ to an even-grade multivector, by introducing bivector and pseudoscalar fields:
\begin{equation*}
f \to \psi_e = f \, 1 + \frac{1}{2!} F_{\mu \nu} \gamma^\mu \wedge \gamma^\nu + g I,
\end{equation*}
each component of which also obeys the Klein-Gordon equation:
\begin{equation*}
\Box \psi_e = -{\omega_0}^2 \psi_e.
\end{equation*}
We now factorize the second-order differential operator $\Box$ into two factors of the first-order operator $\nabla$:
\begin{equation*}
\Box \psi_e = \nabla \nabla \psi_e.
\end{equation*}
With the introduction of a second multivector field $\psi_o$, the Klein-Gordon equation can now be written as a pair of coupled first-order multivector equations:
\begin{align*}
\nabla \psi_e & = \omega_0 \psi_o \\
\nabla \psi_o & = - \omega_0 \psi_e.
\end{align*}
Because $\nabla \psi_e$ is the product of an odd (vector-valued) differential operator with an even multivector field, $\psi_o$ must itself consist only of odd-grade fields,
\begin{equation*}
\psi_o = v_\mu \gamma^\mu + p_\mu I \gamma^\mu.
\end{equation*}
The final form of the multivector Dirac equation will depend upon our assumptions about the relationship between $\psi_o$ to $\psi_e$. If we assume that the vector and pseudovector fields in $\psi_o$ are coupled to, but otherwise independent of, the scalar, bivector, and pseudoscalar fields, then we can combine the two equations into a single complex multivector equation, via the introduction of an imaginary unit scalar:
\begin{equation}
\psi_M = \psi_e + j \psi_o.
\end{equation}
The imaginary unit $j$, unlike the roots of $-1$ in the spacetime algebra, has no geometric interpretation. The spacetime algebra $Cl_{1,3}(\mathbb{R})$ has been complexified to $Cl_{1, 3}(\mathbb{C}) = \mathbb{C} \otimes Cl_{1,3}(\mathbb{R})$, and the Dirac equation takes the form
\begin{equation}
j \nabla \psi_M = \omega_0 \psi_M.
\end{equation}
Breaking these equations up into their components, we acquire the ``massive'' version of Maxwell's equations,
\begin{align*} \label{eq:massivemaxwell}
\partial^\alpha f + \partial_\beta F^{\beta \alpha} & = \omega_0 v^\alpha \\
\partial^\alpha g + \partial_\beta \mathscr{F}^{\beta \alpha} & = - \omega_0 p^\alpha \\
\partial_\alpha v^\alpha & = - \omega_0 f \\
\partial_\alpha p^\alpha & = \omega_0 g \\
F_v^{\alpha \beta} + F_p^{\alpha \beta} & = - \omega_0 F^{\alpha \beta}.
\end{align*}

Conventional applications of the spacetime algebra avoid introducing non-geometric elements such as $j$, seeking instead a geometric element within the real algebra $Cl_{1,3}(\mathbb{R})$ to take its place. In such applications, $\psi_o$ is not independent of $\psi_e$. Rather, it is constructed by right-multiplying $\psi_e$ by some odd basis element $M_-$:
\begin{equation*}
\psi_o = \psi_e M_-.
\end{equation*}
By convention,
\begin{equation*}
M_- = \gamma^2 \gamma^1 \gamma^0,
\end{equation*}
and the Dirac equation in this formulation becomes
\begin{equation}
\nabla \psi_e = \omega_0 \psi_e \gamma^2 \gamma^1 \gamma^0.
\end{equation}
This introduces a preferred direction (or, hypervolume) into the equations of motion. This introduction is justified when $\psi_e$ is interpreted as a transformation operator that acts on the fixed reference bispinor $w$ and basis elements $\gamma^\mu$ to yield dynamic bispinors and multivector quantities, e.g.,
\begin{align*}
w & \to \psi(x) = \psi_M (x) w \\
\gamma^\mu & \to e^\mu (x) = \psi_M (x) \gamma^\mu \wt{\psi}_M (x).
\end{align*}
Quantities such as $\psi(x)$ and $e^\mu (x)$, and not $\psi_e (x)$ itself, are given physical meaning in the operator interpretation, and the presence of fixed multivectors in the Dirac equation is therefore not considered problematic.

\subsection{Introducing the electric charge}
With the introduction of electric charge, again using the algebra $\mathbb{C} \otimes Cl_{1,3}(\mathbb{R})$, the multivector Dirac equation becomes
\begin{equation*}
j \nabla \psi_M - e A \psi_M = \omega_0 \psi_M,
\end{equation*}
where
\begin{equation*}
A \equiv A_\mu \gamma^\mu.
\end{equation*}
The electric charge couples the real and imaginary terms of the same grade, requiring us to introduce a second multivector field:
\begin{align} \label{eq:fullmultivector}
\psi_M \to \psi_1 + j \psi_2 \nonumber \\
\psi_1 = \psi_e + j \psi_o \nonumber \\
\psi_2 = \psi'_e + j \psi'_o.
\end{align}
The Dirac equation now becomes
\begin{align} \label{eq:massivechargedmaxwell}
\nabla \psi_e - e A \psi'_e - \omega_0 \psi_o & = 0 \nonumber \\
\nabla \psi'_e + e A \psi_e - \omega_0 \psi'_o & = 0 \nonumber \\
\nabla \psi_o - e A \psi'_o + \omega_0 \psi_e & = 0 \nonumber \\
\nabla \psi'_o + e A \psi_o + \omega_0 \psi'_e & = 0,
\end{align}
or, in component form,
\begin{align}
(\partial^\alpha f - e A^\alpha f') + (\partial_\beta F^{\beta \alpha} - e A_\beta F'^{\beta \alpha}) & = \omega_0 v^\alpha \nonumber \\
(\partial^\alpha f' + e A^\alpha f) + (\partial_\beta F'^{\beta \alpha} + e A_\beta F^{\beta \alpha}) & = \omega_0 v'^\alpha \nonumber \\
(\partial^\alpha g - e A^\alpha g') + (\partial_\beta \mathscr{F}^{\beta \alpha} - e A_\beta \mathscr{F}'^{\beta \alpha}) & = - \omega_0 p^\alpha \nonumber \\
(\partial^\alpha g' + e A^\alpha g) + (\partial_\beta \mathscr{F}'^{\beta \alpha} + e A_\beta \mathscr{F}^{\beta \alpha}) & = - \omega_0 p'^\alpha \nonumber \\
({F_v}^{\alpha \beta} - e {F_{Av'}}^{\alpha \beta}) + ({F_p}^{\alpha \beta} - e {F_{Ap'}}^{\alpha \beta}) & = - \omega_0 F^{\alpha \beta} \nonumber \\
({F_{v'}}^{\alpha \beta} + e {F_{Av}}^{\alpha \beta}) + ({F_{p'}}^{\alpha \beta} + e {F_{Ap}}^{\alpha \beta}) & = - \omega_0 F'^{\alpha \beta} \nonumber \\
\partial_\alpha v^\alpha - e A_\alpha v'^{\alpha} & = - \omega_0 f \nonumber \\
\partial_\alpha v'^\alpha + e A_\alpha v^\alpha & = - \omega_0 f' \nonumber \\
\partial_\alpha p^\alpha - e A_\alpha p'^{\alpha} & = \omega_0 g \nonumber \\
\partial_\alpha p'^\alpha + e A_\alpha p^\alpha & = \omega_0 g',
\end{align}
where we have defined
\begin{align*}
{F_{Av}}^{\alpha \beta} & \equiv A^\alpha v^\beta - A^\beta v^\alpha \\
{F_{Av'}}^{\alpha \beta} & \equiv A^\alpha v'^\beta - A^\beta v'^\alpha \\
{F_{Ap}}^{\alpha \beta} & \equiv \frac{1}{2!} \epsilon^{\alpha \beta \gamma \delta} (A_\gamma p_\delta - A_\delta p_\gamma) \\
{F_{Ap'}}^{\alpha \beta} & \equiv \frac{1}{2!} \epsilon^{\alpha \beta \gamma \delta} (A_\gamma p'_\delta - A_\delta p'_\gamma).
\end{align*}

\section{Discrete symmetries}
The discrete charge conjugation, parity, and time reversal transformations are \cite{itzykson-zuber}:
\begin{align*}
\psi(x) & \xrightarrow{C} i \gamma^2 \psi^*(x) \\
\psi(x) & \xrightarrow{P} \gamma^0 \psi(\overline{x}) \\
\psi(x) & \xrightarrow{T} i \gamma^1 \gamma^3 \psi^*(-\overline{x}),
\end{align*}
where
\begin{align*}
x & = (t, \vec{x}) \\
\overline{x} & = (t, -\vec{x}).
\end{align*}
$C$, $P$, and $T$ transform solutions of the Dirac equation into solutions of the Dirac equation with, respectively,
\begin{align*}
e & \xrightarrow{C} -e \\
\vec{x} & \xrightarrow{P} -\vec{x} \\
t, \vec{A} & \xrightarrow{C} -t, -\vec{A}.
\end{align*}
(Under time reversal, we assume that source currents, and therefore the vector potentials, also change sign, in addition to the time coordinate itself.) In the operator interpretation, the corresponding multivector transformations are \cite{doran-lasenby}:
\begin{align*}
\psi_M(x) & \xrightarrow{C} - \psi_M(x) \gamma^1 \gamma^0 \\
\psi_M(x) & \xrightarrow{P} \gamma^0 \psi_M(\overline{x}) \gamma^0 \\
\psi_M(x) & \xrightarrow{T} -I \gamma^0 \psi_M(-\overline{x}) \gamma^1.
\end{align*}
We shall see below how these transformations take on a more physically meaningful form in the field interpretation.

\subsection{Charge conjugation}
Note that changing the relative sign of $\psi_1$ and $\psi_2$ in eq. \ref{eq:fullmultivector} is equivalent to changing the sign of the charge $e$ in eq. \ref{eq:massivechargedmaxwell}. We will see that the charge conjugation operation $\psi \to \gamma^2 \psi^*$ accomplishes this relative sign change.

Under complex conjugation, $j \to -j$ and $\gamma^\mu \to (\gamma^\mu)^*$. In both the Dirac and Weyl basis, $\gamma^2$ is the only gamma matrix with imaginary components, and complex conjugation of the gamma matrices can therefore be written as the transformation
\begin{equation*}
\gamma^\mu \to (\gamma^\mu)^* = \gamma^2 \gamma^\mu \gamma^2.
\end{equation*}
This transformation does not, however, extend to general elements of the algebra. For example, the complex conjugate of a product of two gamma matrices is
\begin{align*}
(\gamma^\mu \gamma^\nu)^* & = (\gamma^\mu)^* (\gamma^\nu)^* \\
& = (\gamma^2 \gamma^\mu \gamma^2) (\gamma^2 \gamma^\nu \gamma^2) \\
& = - \gamma^2 \gamma^\mu \gamma^\nu \gamma^2 \\
& \neq \gamma^2 (\gamma^\mu \gamma^\nu) \gamma^2.
\end{align*}
Rather, the general complex conjugate transformation is:
\begin{equation*}
\psi_M \to \psi^*_M = \gamma^2 I \psi_M I \gamma^2.
\end{equation*}
Because odd grades anticommute with $I$, we recover the previous operation for $\psi_o$ and $\psi'_o$:
\begin{equation*}
\gamma^2 I \psi_o I \gamma^2 = -\gamma^2 I I \psi_o \gamma^2 = \gamma^2 \psi_o \gamma^2.
\end{equation*}
Even grades, on the other hand, commute with $I$, introducing the extra negative sign necessary for $\psi_e$ and $\psi'_e$:
\begin{equation*}
\gamma^2 I \psi_e I \gamma^2 = \gamma^2 I I \psi_e \gamma^2 = - \gamma^2 \psi_e \gamma^2.
\end{equation*}
The complex conjugate $\psi^* = \psi_M^* w$ can now be written
\begin{equation*}
(\gamma^2 I \overline{\psi}_M I \gamma^2) w.
\end{equation*}
Our projection bispinor $w$ is real and therefore unchanged by complex conjugation, and $\psi_M \to \overline{\psi}_M$ changes the sign of all factors of $j$:
\begin{align*}
\overline{\psi}_1 & = \psi_e - j \psi_o \\
\overline{\psi}_2 & = \psi'_e - j \psi'_o \\	
\overline{\psi}_M & = \overline{\psi}_1 - j \overline{\psi}_2.
\end{align*}
The conjugation $\psi_{1,2} \to \overline{\psi}_{1,2}$ changes the sign of the odd elements while leaving even elements unchanged, and is therefore equivalent to the spacetime transformation
\begin{equation*}
\psi_i \to \overline{\psi}_i = -I \psi_i I,
\end{equation*}
allowing us to write
\begin{equation} \label{eq:complexconjugate}
\overline{\psi}_M = -I (\psi_1 - j \psi_2) I = -I \psi^c_M I.
\end{equation}
The charge-conjugate multivector
\begin{equation} \label{eq:conjugatemultivector}
\psi^c_M = \psi_1 - j \psi_2
\end{equation}
contains the sign change necessary to transform a field with charge $e$ into one with charge $-e$.
The full charge conjugation operation $\psi \to \gamma^2 \psi^*$ becomes (ignoring the irrelevant phase factor $i$)
\begin{align*}
\psi_M w & \to \gamma^2 (\gamma^2 I \overline{\psi}_M I \gamma^2) w \\
& = \gamma^2 (\gamma^2 I) (-I \psi^c_M I) (I \gamma^2) w \\
& = \psi^c_M (\gamma^2 w),
\end{align*}
or,
\begin{align}
\psi_M & \xrightarrow{C} \psi^c_M \nonumber \\
w & \xrightarrow{C} \gamma^2 w.
\end{align}

\subsection{Parity reversal}
Under the parity reversal operation, all spatial vectors change sign:
\begin{align*}
\gamma^0 & \to \gamma^0 \\
\gamma^i & \to - \gamma^i,
\end{align*}
which can be written as the transformation
\begin{equation*}
\gamma^\mu \to \gamma^0 \gamma^\mu \gamma^0.
\end{equation*}
This extends to general multivectors as well:
\begin{align*}
\gamma^{\mu_1} \gamma^{\mu_2} \cdots \gamma^{\mu_n} & \to (\gamma^0 \gamma^{\mu_1} \gamma^0) (\gamma^0 \gamma^{\mu_2} \gamma^0) \cdots (\gamma^0 \gamma^{\mu_n} \gamma^0) \\
& = \gamma^0 (\gamma^{\mu_1} \gamma^{\mu_2} \cdots \gamma^{\mu_n}) \gamma^0.
\end{align*}
The Dirac bispinor transforms as
\begin{equation*}
\psi(x) \to \gamma^0 \psi(\overline{x}),
\end{equation*}
which yields the multivector transformation
\begin{equation*}
\psi_M w \to (\gamma^0 \psi_M \gamma^0) (\gamma^0 w),
\end{equation*}
or
\begin{align}
\psi_M & \xrightarrow{P} \gamma^0 \psi_M \gamma^0 \nonumber \\
w & \xrightarrow{P} \gamma^0 w.
\end{align}
The transformed multivector is the expected parity-reversed form of $\psi_M$.

\subsection{Time reversal}
Under the time reversal operation $T$, the timelike basis vector changes sign, and the spacelike basis vectors remain unchanged:
\begin{align*}
\gamma^0 & \to -\gamma^0 \\
\gamma^i & \to \gamma^i.
\end{align*}
The negative of the parity reversal operation,
\begin{equation*}
\gamma^\mu \to - \gamma^0 \gamma^\mu \gamma^0
\end{equation*}
accomplishes this transformation for vectors, but not for the general multivector. As with complex conjugation, we can can resolve this discrepancy with a slight modification:
\begin{equation*}
\gamma^\mu \to - \gamma^0 I \gamma^\mu I \gamma^0.
\end{equation*}
Compare this operation to the conventional time reversal operation, which is (up to a phase):
\begin{equation*}
\psi(x) \to \gamma^1 \gamma^3 \psi^*(-\overline{x}).
\end{equation*}
The factorized field transforms as:
\begin{align*}
\psi_M w & \to \gamma^1 \gamma^3 (\gamma^2 I \overline{\psi}_M I \gamma^2) w \\
& = \gamma^1 \gamma^3 \gamma^2 I (-I \psi^c_M I) I \gamma^2 w \\
& = (- \gamma^0 I \psi^c_M I \gamma^0) (I \gamma^0 \gamma^2 w),
\end{align*}
or
\begin{align}
\psi_M & \xrightarrow{T} -\gamma^0 I \psi^c_M I \gamma^0 \nonumber \\
w & \xrightarrow{T} I \gamma^0 \gamma^2 w.
\end{align}
That is, $\psi_M$ undergoes a time reversal,
\begin{equation*}
\psi_M \to -\gamma^0 I \psi_M I \gamma^0,
\end{equation*}
in addition to a charge conjugation,
\begin{equation*}
\psi_M \to \psi^c_M.
\end{equation*}
The combined time reversal and charge conjugation gives the expected transformation for the electromagnetic potential:
\begin{align*}
q (\gamma^0 \phi + \gamma^i A_i) & \xrightarrow{T} q (-\gamma^0 \phi + \gamma^i A_i) \\
& \xrightarrow{C} q (\gamma^0 \phi - \gamma^i A_i).
\end{align*}

\subsection{CPT}
In summary, charge conjugation changes the relative sign between the two multivector fields $\psi_1$ and $\psi_2$, which are coupled by the electric charge. This effectively changes the sign of the charge for the full multivector field $\psi_M$:
\begin{equation*}
\psi_M = \psi_1 + j \psi_2 \xrightarrow{C} \psi^c_M = \psi_1 - j \psi_2.
\end{equation*}
Parity reversal changes the sign of all spatial vectors, and time reversal changes the sign of all time vectors in addition to performing a charge conjugation. The net effect under a combined $CPT$ transformation is
\begin{align*}
\psi_M(x) & \xrightarrow{C} \psi^c_M(x) \\
& \xrightarrow{P} \gamma^0 \psi^c_M(\overline{x}) \gamma^0 \\
& \xrightarrow{T} -I \psi_M(-x) I.
\end{align*}
The two charge conjugation operations cancel, and the multivector field undergoes a full spacetime reversal.

\section{A real representation of the Dirac algebra}
In our field-based interpretation, the imaginary unit scalar $j$ is not a part of the real spacetime algebra $Cl_{1,3}(\mathbb{R})$. For the massive Dirac equation, we have a complex-valued spacetime algebra:
\begin{equation*}
Cl_{1,3}(\mathbb{C}) = \mathbb{C} \otimes Cl_{1,3}(\mathbb{R}).
\end{equation*}
In the operator interpretation, which utilizes the real spacetime algebra
\begin{equation*}
Cl_{1,3}(\mathbb{R}),
\end{equation*}
there is no imaginary unit scalar element that commutes with all other elements, but one can be implemented by taking advantage of the projective nature of $w$:
\begin{align*}
\begin{pmatrix} i \boldsymbol{1} & \boldsymbol{0} \\ \boldsymbol{0} & i \boldsymbol{1} \end{pmatrix} w & = \begin{pmatrix} i \boldsymbol{\sigma}^3 & \boldsymbol{0} \\ \boldsymbol{0} & i \boldsymbol{\sigma}^3 \end{pmatrix} w \\
\to i w & = \gamma^2 \gamma^1 w.
\end{align*}
In the field interpretation, we wish to distinguish this geometric implementation from the non-geometric $j$. To that end, in this section we develop a real orthogonal representation for the Dirac algebra.
 
The real spacetime algebra is isomorphic to the algebra of $2 \times 2$ quaternion-valued matrices:
\begin{equation*}
Cl_{1,3}(\mathbb{R}) \cong M_2(\mathbb{H}) = M_2(\mathbb{R}) \otimes \mathbb{H}.
\end{equation*}
The algebra of $2 \times 2$ real-valued matrices, $M_2(\mathbb{R})$, contains four basis elements,
\begin{equation*}
\{1, M_1, M_2, M_3 \},
\end{equation*}
where $1$ is the identity element and
\begin{align*}
M_1^2 & = -M_2^2 = M_3^2 = 1 \\
M_1 M_2 & = - M_2 M_1 = M_3 \\
M_2 M_3 & = - M_3 M_2 = M_1 \\
M_3 M_1 & = - M_1 M_3 = - M_2.
\end{align*}
The quaternion algebra $\mathbb{H}$ also contains four basis elements,
\begin{equation*}
\{1, I_1, I_2, I_3 \},
\end{equation*}
with identity element $1$ and
\begin{equation*}
I_i I_j = - \delta_{i j} \, 1 + \epsilon_{i j k} I_k.
\end{equation*}
The product algebra can now be constructed:
\begin{align*}
1 & \mapsto 1 \otimes 1 \\
\gamma^0 & \mapsto M_3 \otimes 1 \\
\gamma^i & \mapsto M_1 \otimes I_i \\
\gamma^i \wedge \gamma^0 & \mapsto M_2 \otimes I_i \\
\gamma^i \wedge \gamma^j & \mapsto \epsilon_{i j k} 1 \otimes I_k \\
I \gamma^0 & \mapsto -M_1 \otimes 1 \\
I \gamma^i & \mapsto M_3 \otimes I_i \\
I & \mapsto -M_2 \otimes 1.
\end{align*}

In the conventional Dirac basis, both $M_2(\mathbb{R})$ and $\mathbb{H}$ are represented as $2 \times 2$ matrices with complex-valued components:
\begin{align*}
M_1 & \mapsto \begin{pmatrix} 0 & i \\ -i & 0 \end{pmatrix} \\
M_2 & \mapsto \begin{pmatrix} 0 & -i \\ -i & 0 \end{pmatrix} \\
M_3 & \mapsto \begin{pmatrix} 1 & 0 \\ 0 & -1 \end{pmatrix} \\
I_1 & \mapsto \begin{pmatrix} 0 & -i \\ -i & 0 \end{pmatrix} \\
I_2 & \mapsto \begin{pmatrix} 0 & -1 \\ 1 & 0 \end{pmatrix} \\
I_3 & \mapsto \begin{pmatrix} -i & 0 \\ 0 & i \end{pmatrix}.
\end{align*}
For our purposes, however, we wish to choose real orthogonal matrices to represent each algebra $\mathbb{C}$, $M_2(\mathbb{R})$, and $\mathbb{H}$. Let us first represent $\mathbb{C}$ as $2 \times 2$ matrices:
\begin{align*}
1 & \mapsto \begin{pmatrix} 1 & 0 \\ 0 & 1 \end{pmatrix} \\
j & \mapsto \begin{pmatrix} 0 & -1 \\ 1 & 0 \end{pmatrix},
\end{align*}
with basis spinors
\begin{equation*}
a_1 = \begin{pmatrix} 1 \\ 0 \end{pmatrix} \qquad a_2 = \begin{pmatrix} 0 \\ 1 \end{pmatrix}.
\end{equation*}
The $2 \times 2$ real matrices in $M_2(\mathbb{R})$ are represented as:
\begin{align*}
1 & \mapsto \begin{pmatrix} 1 & 0 \\ 0 & 1 \end{pmatrix} \\
M_1 & \mapsto \begin{pmatrix} 0 & 1 \\ 1 & 0 \end{pmatrix} \\
M_2 & \mapsto \begin{pmatrix} 0 & -1 \\ 1 & 0 \end{pmatrix} \\
M_3 & \mapsto \begin{pmatrix} 1 & 0 \\ 0 & -1 \end{pmatrix}
\end{align*}
with basis spinors
\begin{equation*}
b_1 = \begin{pmatrix} 1 \\ 0 \end{pmatrix} \qquad b_2 = \begin{pmatrix} 0 \\ 1 \end{pmatrix}.
\end{equation*}
Finally, we represent the quaternion algebra $\mathbb{H}$ using real $4 \times 4$ matrices, by effectively taking the complex $2 \times 2$ representation and then representing $0$, $1$, and $i$ by their corresponding $2 \times 2$ matrices in the above representation of $\mathbb{C}$:
\begin{align*}
1 & \mapsto \begin{pmatrix} 1 & 0 & 0 & 0 \\ 0 & 1 & 0 & 0 \\ 0 & 0 & 1 & 0 \\ 0 & 0 & 0 & 1 \end{pmatrix} \\
I_1 & \mapsto \begin{pmatrix} 0 & 0 & 0 & 1 \\ 0 & 0 & -1 & 0 \\ 0 & 1 & 0 & 0 \\ -1 & 0 & 0 & 0 \end{pmatrix} \\
I_2 & \mapsto \begin{pmatrix} 0 & 0 & -1 & 0 \\ 0 & 0 & 0 & -1 \\ 1 & 0 & 0 & 0 \\ 0 & 1 & 0 & 0 \end{pmatrix} \\
I_3 & \mapsto \begin{pmatrix} 0 & 1 & 0 & 0 \\ -1 & 0 & 0 & 0 \\ 0 & 0 & 0 & -1 \\ 0 & 0 & 1 & 0 \end{pmatrix},
\end{align*}
with four basis spinors
\begin{equation*}
c_1 = \begin{pmatrix} 1 \\ 0 \\ 0 \\ 0 \end{pmatrix} \qquad c_2 = \begin{pmatrix} 0 \\ 1 \\ 0 \\ 0 \end{pmatrix} \qquad c_3 = \begin{pmatrix} 0 \\ 0 \\ 1 \\ 0 \end{pmatrix} \qquad c_4 = \begin{pmatrix} 0 \\ 0 \\ 0 \\ 1 \end{pmatrix}.
\end{equation*}

With these representations, we can now construct a projection bispinor for the full algebra $\mathbb{C} \otimes M_2(\mathbb{R}) \otimes \mathbb{H}$ as a tensor product of the first basis spinors from each algebra:
\begin{equation*}
w = a_1 \otimes b_1 \otimes c_1
\end{equation*}
Bilinears formed using this projection bispinor will isolate the $(1,1)$ component of each algebra, i.e., the real component from $\mathbb{C}$ and the $1$ or $\gamma^0$ components from $Cl_{1,3}(\mathbb{R})$:
\begin{equation}
w^T M w = \text{Re} \langle M ( 1 + \gamma^0 ) \rangle.
\end{equation}
For example, the vector bilinear for the uncharged massive Dirac field is:
\begin{align*}
\overline{\psi} \gamma^\mu \psi & = w^T \gamma^0 (\wt{\psi}_e - j \wt{\psi}_o) \gamma^\mu (\psi_e + j \psi_o) w \\
& = \text{Re} \langle \gamma^0 ( \wt{\psi}_e - j \wt{\psi}_o) \gamma^\mu (\psi_e + j \psi_o ) (1 + \gamma^0) \rangle \\
& = \langle (\psi_e \gamma^0 \wt{\psi}_e + \psi_o \gamma^0 \wt{\psi}_o) \gamma^\mu \rangle,
\end{align*}
which naturally extends the vector bilinear of the massless Dirac field to include odd field components:
\begin{equation*}
j = \psi_e \gamma^0 \wt{\psi}_e + \psi_o \gamma^0 \wt{\psi}_o.
\end{equation*}
Likewise, the spin bilinear
\begin{equation*}
S^{0 i j} = \frac{i}{2} \overline{\psi} \gamma^0 (\gamma^i \wedge \gamma^j) \psi,
\end{equation*}
with $i w \to \gamma^2 \gamma^1 w$, becomes
\begin{equation*}
S^{0 i j} = \frac{1}{2} \langle (\psi_e \gamma^2 \gamma^1 \gamma^0 \wt{\psi}_e + \psi_o \gamma^2 \gamma^1 \gamma^0 \wt{\psi}_o) \gamma^0 (\gamma^i \wedge \gamma^j) \rangle,
\end{equation*}
a straightforward extension of the massless field's spin bilinear. We could equally well, however, implement $i$ using our non-geometric unit scalar $j$, resulting in a distinct pseudovector quantity, which mixes even and odd grades and does not use the reference pseudovector $\gamma^2 \gamma^1 \gamma^0$:
\begin{equation*}
S^{0 i j} = \frac{1}{2} \langle (\psi_e \wt{\psi}_o - \psi_o \wt{\psi}_e) \gamma^0 (\gamma^i \wedge \gamma^j) \rangle.
\end{equation*}
The two pseudovectors coincide (up to a sign) if we impose the relation $\psi_o = \psi_e \gamma^2 \gamma^1 \gamma^0$, as is the case in the operator interpretation. In the field interpretation, however, we see that the door is open for alternate constructions of the angular momentum, the Lagrangian, and other such bilinear quantities.

\bibliography{on-the-physical-interpretation-of-the-dirac-wavefunction-ii-the-massive-dirac-field}
\bibliographystyle{hunsrt}

\end{document}